\documentclass[journal]{IEEEtran}

\usepackage{cite}
\usepackage[pdftex]{graphicx}
\graphicspath{{./images/}}
\usepackage{makecell}
\usepackage{amsmath}
\usepackage{algorithmic}
\usepackage{array}
\usepackage{fixltx2e}
\usepackage{dblfloatfix}
\usepackage{url}
\usepackage{siunitx}
\usepackage[ruled,vlined]{algorithm2e}
\usepackage{mathtools, nccmath}
\DeclarePairedDelimiter{\nint}\lfloor\rceil

\begin{document}

\title{Automated Environmental Compliance Monitoring with IoT and Open Government Data}

\author{Lizaveta~Miasayedava,~\IEEEmembership{Graduate~Student~Member,~IEEE,}
        Keegan~McBride,
        and~Jeffrey~Andrew~Tuhtan,~\IEEEmembership{Member,~IEEE}%
\thanks{L. Miasayedava is with the Research Laboratory for Proactive Technologies, Tallinn University of Technology, Tallinn 12616, Estonia (email: lizaveta.miasayedava@taltech.ee).}%
\thanks{K. McBride is with Hertie School's Centre for Digital Governance, Berlin, Germany (email: mcbride@hertie-school.org).}%
\thanks{J. A. Tuhtan is with the Centre for Biorobotics, Tallinn University of Technology, Tallinn 12616, Estonia (email: jeffrey.tuhtan@taltech.ee).}%
}

\markboth{SUBMITTED TO IEEE INTERNET OF THINGS JOURNAL}%
{Miasayedava \MakeLowercase{\textit{et al.}}: Automated Environmental Compliance Monitoring with IoT and Open Government Data}

\maketitle
\begin{abstract}
Negative environmental impacts on societies and ecosystems are frequently driven by human activity and amplified by increasing climatic variability. Properly managing these impacts relies on a government's ability to ensure environmental regulatory compliance in the face of increasing uncertainty. Water flow rates are the most widely used evaluation metric for river regulatory compliance. Specifically, compliance thresholds are set by calculating the minimum flow rates required by aquatic species such as fish. These are then designated as the minimum "environmental flows" (eflows) for each river. In this paper, we explore how IoT-generated open government data can be used to enhance the development of an automated IoT-based eflows compliance system. To reduce development and operational costs, the proposed solution relies on routinely collected river monitoring data. Our approach allows for any authority with similar data to rapidly develop, test and verify a scalable solution for eflow regulatory compliance monitoring and evaluation. Furthermore, we demonstrate a real-world application of our system using open government data from Estonia's national river monitoring network. The main novelty of this work is that the proposed IoT-based system provides a simple evaluation tool that re-purposes IoT-generated open government data to evaluate compliance and improve monitoring at a national scale. This work showcases a new paradigm of IoT-based solutions using open government data and provides a real-world example of how the solution can automatically evaluate environmental compliance in increasingly uncertain environments.
\end{abstract}


\IEEEpeerreviewmaketitle

\section{Introduction}

\begin{figure*}[!t]
\centering
\includegraphics[width=\textwidth]{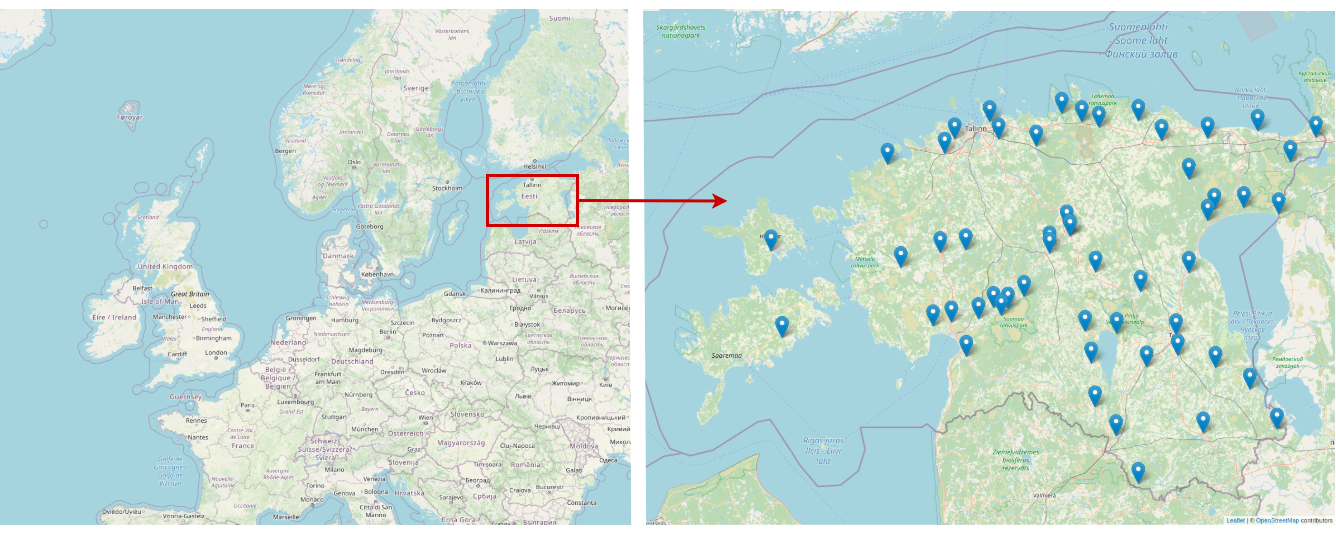}
\caption{Map showing the locations of the 54 sites which make up the Estonian river hydrological monitoring network, the map layer is retrieved from \cite{OpenStreetMap}, data from the sites were taken from the open government web portal \cite{weather-service} from 2009 to 2018.}
\label{fig_map}
\end{figure*}

\IEEEPARstart{C}{limate} change is a major threat to the stability of human society and river ecosystems \cite{vorosmarty2010}. Extreme events such as droughts and floods can have significant economic impacts across multiple sectors, including agriculture, hydropower and inland shipping \cite{ludwig2012}. These extreme events also act as stressors to aquatic ecosystems, which are now facing an unprecedented decline in biodiversity \cite{vaughn2010}. As these events are increasing in both frequency and severity, governments must learn how to use the vast amounts of data they already possess to improve societal resilience. To achieve this, new Internet of Things (IoT)-based methods are needed as they can improve monitoring, prediction and response to extreme events. This paper explores how IoT-generated open government data (OGD) can be leveraged to automate environmental compliance in a large-scale river sensing network, as shown in Fig. \ref{fig_map}.

Governments are tasked with monitoring and addressing potential compound sociological impacts, e.g. via the monitoring of riverine resources or enforcing regulations, but, regulatory agencies are currently faced with challenges that are preventing them from adequately conducting their assigned tasks and required activities \cite{eflows-difficulties}. One such problem is related to the acquisition and usage of high quality, up-to-date, and relevant data; these problems may emerge for both technological (e.g. missing equipment, failing sensors, or underfunded infrastructure) and regulatory (e.g. lacking or missing policy) reasons. One possible way to overcome these challenges and risks is through the use of a combination of IoT networks and OGD. We call this new paradigm the "The Internet of Open Government Data and Things" (IoOGDT), a simple concept which has not been widely studied. Specifically, IoOGDT can provide new and innovative services to assist the government with environmental compliance monitoring, promoting adaptive regulation. This paper presents a large-scale example by showcasing the applicability of IoOGDT to environmental regulatory compliance in rivers within the European Union (EU).

In the EU, the main regulatory objective associated with river ecosystems is to achieve and maintain a high-level ecological status of rivers, as defined in the EU Water Framework Directive (WFD) \cite{wfd}. The WFD serves as the fundamental framework for European water legislation. One aspect specifically set out in the implementation of the WFD is related to the definition, assessment, and monitoring of environmental flows ("eflows"). The definitions of eflows as well as methodologies to estimate them vary \cite{methodologies}, but in general, the concept of eflows can be understood as an indicator of a river’s ecological status, the reporting of which is mandated in the river basin management plans of all EU member states. Moreover, the application of selected methodologies implies various challenges \cite{wfd-clarification}.

Since there is no universally adopted methodology, we implement a simple approach for determining the minimum eflow set forth in the Estonian environmental legislation \cite{estonian-legislation}, automate the assessment at a national scale and subsequently test and validate this approach by using an IoOGDT solution drawing on a network of river discharge, water temperature and water level sensors in Estonia. The design and development of the IoOGDT system followed the Environmental Intelligence (EI) cycle \cite{eic-arctic} - an iterative process of environmental knowledge extraction and application. 

\begin{figure*}[!t]
\centering
\includegraphics[width=\textwidth]{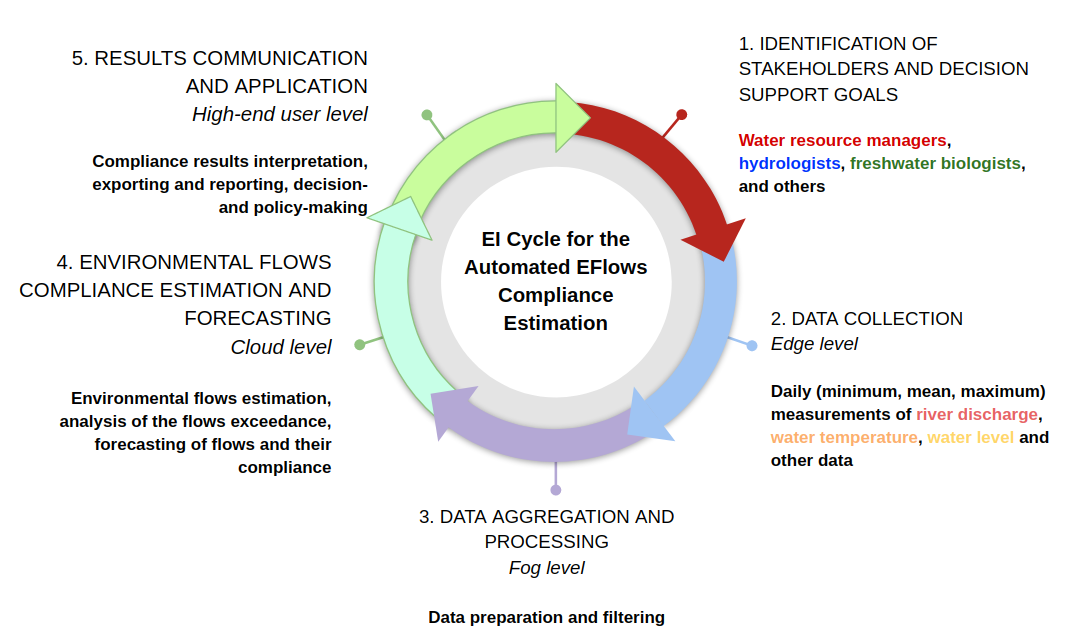}
\caption{Environmental Intelligence (EI) cycle for the automated environmental flows (eflows) compliance estimation.}
\label{fig_eic}
\end{figure*}

\begin{figure*}[!t]
\centering
\includegraphics[width=\textwidth]{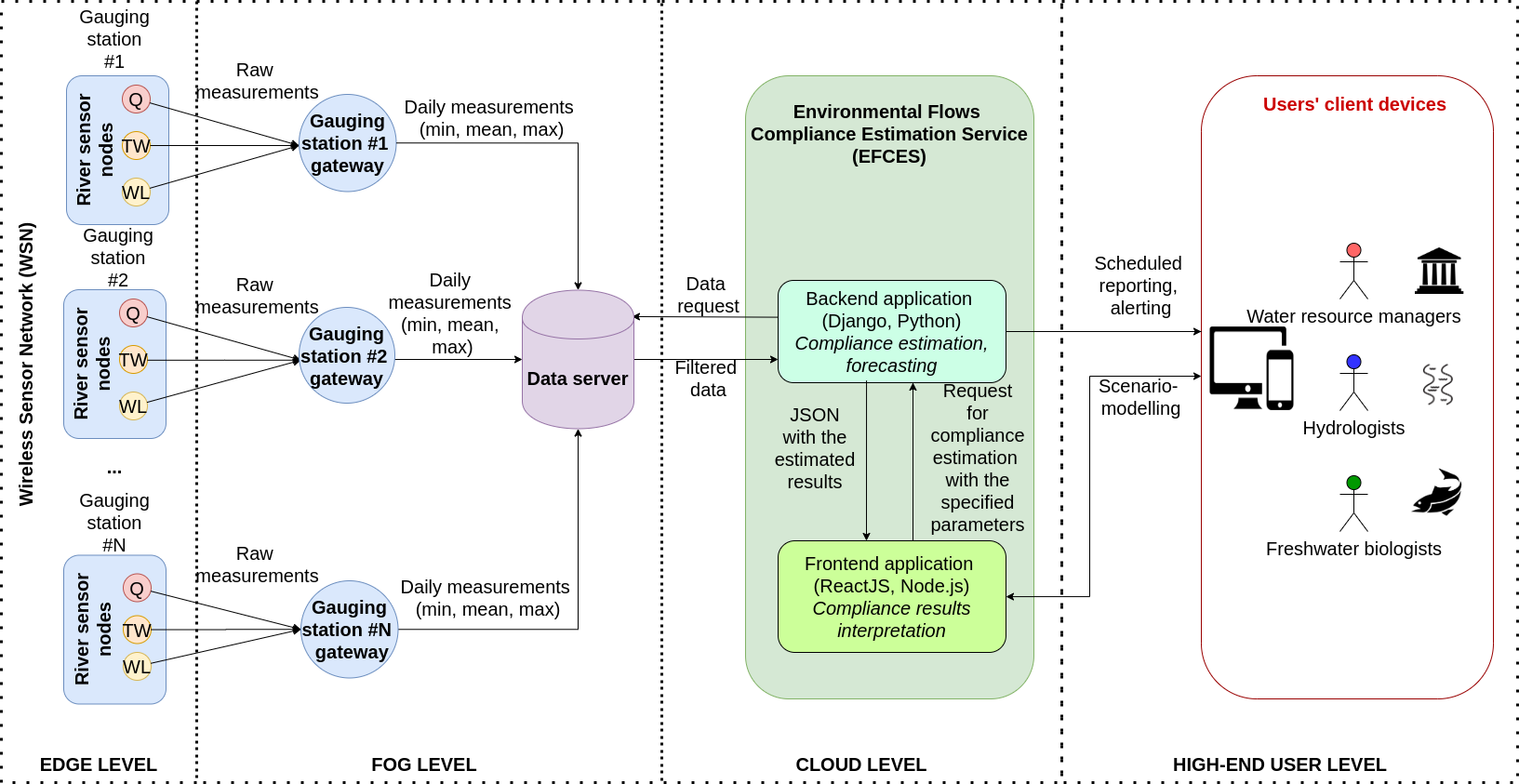}
\caption{Architecture of the proposed real-time environmental flows compliance monitoring system. The edge layer consists of the river sensor nodes which measure the flow rate (Q), water temperature (TW) and water level (WL). The raw measurements undergo data quality control and post-processing, and the daily statistics are pushed to the data server. The compliance estimation service runs at the cloud level.}
\label{fig_architecture}
\end{figure*}

The adaptation of the EI cycle for the eflows compliance estimation is presented in Fig. \ref{fig_eic}. Following this approach, we demonstrate how existing OGD can be transformed into a new innovative service that facilitates improved decision- and policy-making and assessment of regulatory compliance. Embedding the service on the cloud level of an IoT system (see Fig. \ref{fig_architecture}) will help to ensure real-time environmental regulatory compliance monitoring and evaluation.

This article is organised as follows: Section II describes the data sources and IoOGDT-based eflow methodology. Section III presents the architecture and application of the eflow compliance estimation system using the Estonian national hydrological OGD. In section IV, we discuss the strengths, weaknesses and future outlook of IoOGDT for automated environmental regulatory compliance reporting and proactive decision-making support. Finally, Section V contains the conclusions and future research directions.

\section{Related work}

\subsection{Open Government Data and Internet of Things}

A large amount of literature has been devoted to the study of how open government data (OGD) can be used to drive the creation of new innovative services \cite{janssen2012benefits}, \cite{mcbride2019does}, \cite{jetzek2014data}. However, there is a lack of studies that specifically examine OGD that have been created or originate from the Internet of Things (IoT) networks and how they may be utilized. The papers that do cover this topic are almost explicitly related to the topic of smart cities, see, for example \cite{aguilera2017citizen}, \cite{aguilera2016collaboration}, \cite{zanella2014internet}, \cite{ahlgren2016internet}. As previously mentioned, this paper proposes the concept of the Internet of Open Government Data and Things (IoOGDT), which can be understood as data that have been collected or paid for by a government organization, are generated via IoT and subsequently released as OGD (with an appropriate license, in a machine-readable, human-understandable, and freely reusable format). Data released in this way may help to improve and extend the effectiveness and value of the IoT system since anyone can take advantage of them to build a new useful service \cite{mergel2018citizen}.

As OGD may help with improving the quality, relevance, and impact of IoT, so IoT may also help drive the improvement and use of OGD. The primary reason for this is that data coming via an IoOGDT system are expected to be more relevant, accurate, timely, and usable than other standard sources of OGD. This is an important point, because it is known that one of the primary reasons that individuals or citizens do not take advantage of OGD is due to lack of quality, relevance, or timeliness \cite{young2017civic}, \cite{zuiderwijk2012socio}. It is also known that one of the highest predictors for future use of OGD is related to a positive previous experience with an OGD-based system \cite{mcbride2019turning}. Thus, it can be argued that OGD and IoT have a sort of symbiotic relationship, with each improving the other, and that by releasing OGD collected via IoT, it is possible to encourage the creation of new services thereby simultaneously extending and improving the value of the IoT system.

\subsection{Hydrological data}

Considering environmental regulatory compliance in rivers, the most valuable variables come from hydrological data which include the flow rates, water depths and water temperatures. Indeed, the solution proposed in this paper relies on these types of open river data. Because rivers provide a variety of uses including shipping, drinking water, and irrigation, river monitoring data are widely available from local, regional, national and international institutions, and are supported by domain experts with multi-generational knowledge. IoT-based river monitoring systems are now rapidly advancing beyond data logging services to provide data-driven solutions for pumping station control \cite{IoT_Rivers}, increase the automation level of water-intensive agriculture \cite{IoT_Ag} and flood alter systems \cite{IoT_Flood}. Thus, we chose river monitoring as a classical governmental monitoring system to demonstrate how automated compliance can be achieved in a scalable and practical manner by utilizing IoT and simultaneously taking advantage of existing data. In order to demonstrate the proposed solution and to estimate the compliance of river flows, we use Estonian hydrological OGD - publicly accessible Estonian river data collected by the government for weather reporting and water resource management \cite{weather-service}. The OGD used in this work was taken from the Estonian national hydrological network which collects measurements at 54 river gauging stations (see Fig. \ref{fig_map}) from 2009-2018. The hydrological measurements include the daily minimum, average and maximum values of water level (WL) [$cm$], water temperature (TW) [$\si{\degree}C$], and discharge (Q) [$m^3/s$].

\subsection{Compliance estimation methodology}

The compliance estimation of river flows is carried out through the evaluation of the previously described eflows. The cornerstone of eflows compliance estimation is in the choice of the methodology. There are more than 200 conventional methodologies to assess eflows: hydrological, hydraulic-habitat, holistic methodologies, and their combinations \cite{methodologies}. These methods vary in the complexity, assessed characteristics, tools and data requirements and can be used in different situations determining the water use. 

One of the most widely used methods are hydrological methods, due to their ease of use and relatively low cost. For example, in Estonia, according to the regulation of the Ministry of the Environment specifying requirements for the expansion of a water body, environmental monitoring related to the expansion, protection of aquatic life, dam, elimination of the expansion and lowering of the water level, and methodology for determining the minimum ecological flow \cite{estonian-legislation}, the minimum eflow is determined for the ice-free period from May to October by calculating the average monthly minimum flow with a 95\% probability of being exceeded (1).

\begin{equation}
q_{env}=q_{desc}\nint{\frac{p\% * (N+1)}{100\%}},
\end{equation}
where \\
$q_{env}$ - eflow discharge (or volumetric flow rate) [$m^{3}/s$], \\
$q_{desc}$ - array of observed discharge values sorted in the descending order [array of $m^3/s$], \\
$p\%$ - cumulative probability of occurrence for an observed discharge value [\%], \\
$N$ - total count of observations in ${q_{desc}}$. \\

For the solution created in this paper and demonstration of the applicability of the system, formula (1) is used for eflows calculations. This formula requires only river discharge data (Q). However, it is worth noting that the proposed architecture is flexible enough to implement multiple eflow calculation methods.

\section{Proposed system}

\subsection{System architecture}

This section introduces the architecture of the designed IoOGDT-based system for real-time eflows compliance monitoring and evaluation, as well as the developed cloud level of the system - Environmental Flows Compliance Estimation Service (EFCES) enabling eflows compliance estimation and reporting. The code repository is available open-source on GitHub, see \cite{eeflows-repo}.

By design, the system implements the EI cycle, and the use of IoT technologies while implementing the EI cycle may well improve the data collection and aggregation processes, thereby enabling real-time distributed environmental sensing with a network of sensors. This real-time environmental compliance evaluation and monitoring improves the responsiveness to stakeholders’ needs (e.g. the needs of water resource managers, hydrologists, freshwater biologists, government officials, international regulators, and other interested stakeholders), thus ensuring increased awareness about the state of a river ecosystem at a given time.

The edge level of the IoT system is represented by a Wireless Sensor Network (WSN) carrying out data collection. Sensors are located at each of the gauging stations (see Fig. \ref{fig_map}) and collect various hydrological measurements, such as discharge (Q), water temperature (TW), and water level (WL) - hydrological OGD mentioned in Section II B. 

The fog level involves data aggregation and processing, which is organized through passing the collected data to the data server through sink nodes - gateways. The collected daily minimum, average, and maximum measurement data are stored in the data server. 

The cloud level is represented by the previously mentioned and developed EFCES, which accesses the storage to gather the collected data and estimate eflows and their compliance at a particular location (gauging station monitoring a specific river), thus implementing step 4 of the EI cycle. 

EFCES is a web service that implements the following functional requirements:

\begin{enumerate}
    \item Hydrological data acquisition and manipulation.
    \item Environmental flows estimation.
    \item Compliance estimation and interpretation.
    \item Configuration of estimation parameters.
    \item Export of the generated compliance information.
\end{enumerate}

It allows setting the compliance estimation parameters (gauging stations, dates, methods to be used, and other parameters) and simulating various scenarios to adjust the compliance estimation methodology for a river to be used for regulation. The backend application of the service acquires the data from cloud data storage, processes the hydrological data, estimates eflows, and passes the estimated results to the frontend application. The displayed compliance results can then be interpreted, viewed, and exported for reporting as well as for being used in supporting decision- and policy-making ("Results communication and application" step 5 of the EI cycle), thus representing a benefit to water managers who can simply use the tool to generate needed estimations, images, and reports.

On the high-end of the system, water resource managers, hydrologists, freshwater biologists, and other users use the system through the EFCES interface to access reports on the environmental compliance of rivers as well as test various methods and parameters of compliance estimation using the available river data.

\begin{figure*}[!t]
\includegraphics[width=\textwidth]{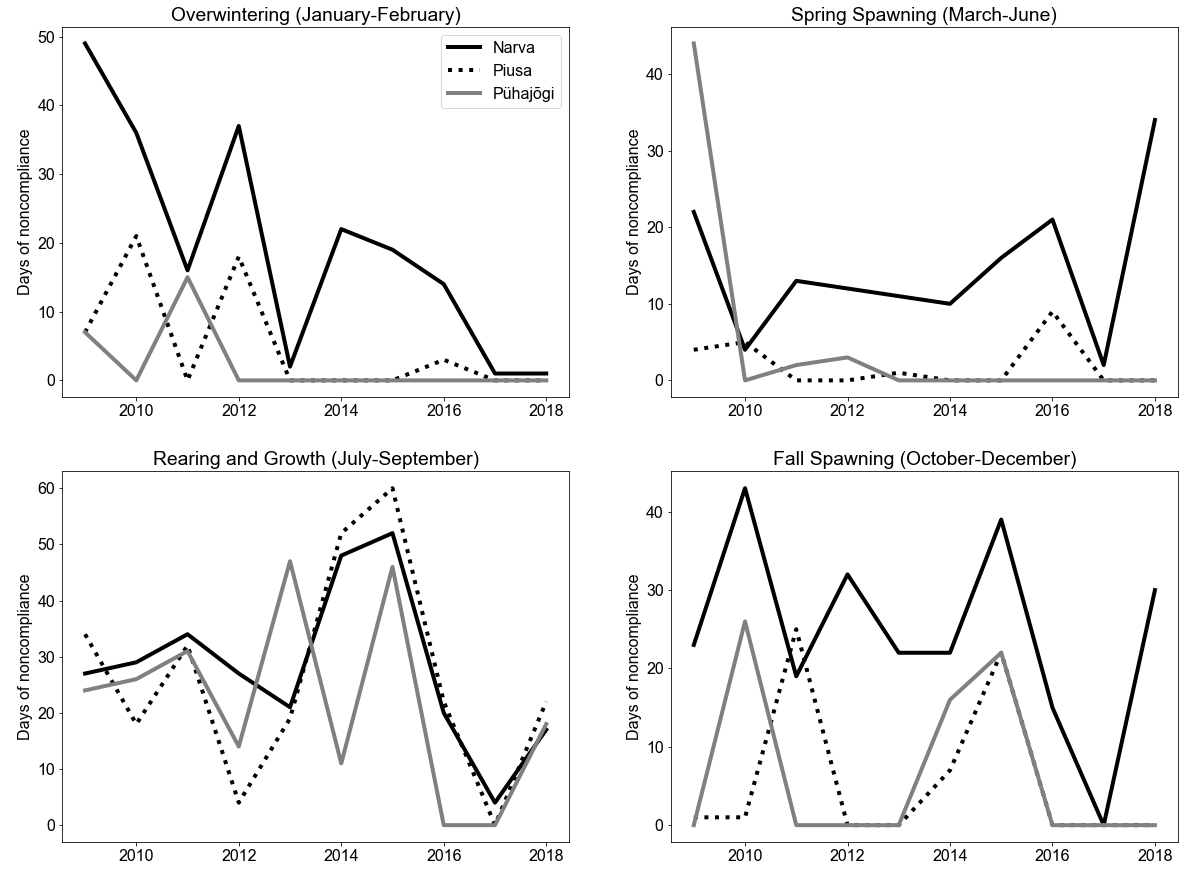}
\caption{Compliance estimation per bioperiod results for rivers monitored at three Estonian gauging stations in 2009-2018. Sorted by their mean annual discharges, the "large" Estonian river Narva is shown with solid black lines, the "medium" river Piusa as dotted black lines, and the "small" Pühajõgi river as solid grey lines.}
\label{fig_plots}
\end{figure*}

\subsection{Compliance estimation demonstration}

For decision- and policy-making as well as monitoring and long-term planning, water managers need to define both the duration and recurrence frequency of noncompliance. In this article, we present the biological approach and calculate the number of all the occurred low flow events (flows below an eflow value calculated according to formula (1)) - the total number of days of noncompliance per each bioperiod over 2009-2018. The months for bioperiods are chosen the same as the ones described in the Polish methodology \cite{raelff}, and can be configured via the user interface of the service.

\begin{table*}[!t]
\renewcommand{\arraystretch}{1.3}
\caption{Eflows Compliance Summary for Three Estonian Rivers in 2009-2018.}
\label{table_01}
\centering
\begin{tabular*}{0.855\textwidth}{|c|c|c|c|c|c|}
  \hline
  River&Minimum discharge [$m^{3}/s$]& 
  \multicolumn{4}{c|}{Duration of noncompliance per bioperiod $[days]$} \\ \cline{3-6}
  &&Overwintering & Spring Spawning & Rearing and Growth & Fall Spawning \\
  \hline 
  Narva (large)
  &\makecell{399.85 $\pm$ 150.32 \\ (132.99; 1210.50)}
  &\makecell{19.70 $\pm$ 15.77 \\ (1; 49)}
  &\makecell{14.50 $\pm$ 8.88 \\ (2; 34)}
  &\makecell{27.90 $\pm$ 13.51 \\ (4; 52)}
  &\makecell{24.50 $\pm$ 11.72 \\ (0; 43)} \\
  \hline
  Piusa (medium)
  &\makecell{5.20 $\pm$ 3.52 \\ (2.45; 43.75)}
  &\makecell{4.90 $\pm$ 7.63 \\ (0; 21)}
  &\makecell{1.90 $\pm$ 2.95 \\ (0; 9)}
  &\makecell{26.30 $\pm$ 17.99 \\ (0; 60)}
  &\makecell{5.60 $\pm$ 9.20 \\ (0; 25)} \\
  \hline
  Pühajõgi (small)
  &\makecell{1.61 $\pm$ 2.04 \\ (0.04; 22.23)}
  &\makecell{2.20 $\pm$ 4.75 \\ (0; 15)}
  &\makecell{4.90 $\pm$ 13.07 \\ (0; 44)}
  &\makecell{21.70 $\pm$ 15.72 \\ (0; 47)}
  &\makecell{6.40 $\pm$ 10.03 \\ (0; 26)} \\
  \hline
\end{tabular*}
\end{table*}

The results of the noncompliance of Estonian rivers were tested in detail at the following three gauging stations: Narva linn (Narva river), Korela (Piusa river), and Toila-Oru (Pühajõgi river) located in eastern Estonia. The corresponding rivers were chosen in order to evaluate the system's ability to estimate noncompliance over all four bioperiods for rivers of different sizes based on their mean annual flow rate values. The days of noncompliance are plotted in Fig. \ref{fig_plots} where the the "large" Narva river (corresponding to the 100th percentile as the largest river in Estonia) is shown as a solid black line, black dotted lines correspond to the "medium" Piusa river (57th percentile) and the "small" river, Pühajõgi (20th percentile) is shown as a grey solid line. In addition to the graphical overview, the summary eflows noncompliance statistics for each of the three selected rivers in 2009-2018 are provided in Table \ref{table_01}. The data are reported as the mean values of each parameter $\pm$ their standard deviation (SD) followed by the range (minimum; maximum) in parentheses.

The results for the first "overwintering" bioperiod from January to February is shown in the top left panel of Fig. \ref{fig_plots}. During this bioperiod, the decreasing trend observed in Narva river is assumed to be related to climate change as there have been no major anthropogenic changes in the river during the period of observation. However, it can be seen that none of the other rivers show a substantial change in the days of noncompliance. Therefore, decision-makers may wish to carry out targeted mitigation measures. This could be accomplished for example by routing more water flows to Narva river, reducing the number of days of noncompliance during the overwintering bioperiod.
 
The "spring spawning" bioperiod runs from March to June, and is shown in the top left panel of Fig. \ref{fig_plots}. Here, the effects of increased winter seasonal air temperature and precipitation in Estonia have resulted in an earlier beginning and decrease in spring floods, and consequently longer dry periods. Again, the most impacted river is Narva river, which clearly shows a recent jump in the days of noncompliance from 2017. Water managers may therefore wish to target this region of the river for mitigation measures, for example by artificially creating more fish spawning habitats to help reduce the stress placed on fish due to the increase in days of eflow noncompliance in this bioperiod. 

The third bioperiod occurs during summer, and is referred to as "rearing and growth". Considering all of the three rivers assessed in detail, a clear biannual pattern from 2011 onwards can be seen in the bottom left panel of Fig. \ref{fig_plots}. This pattern has been attributed to climate change, and has been recognized to have a negative effect not only on fish populations, but also on the water quality (chemical status) and aquatic habitats for other animal species as well \cite{climate-change-estonia}. In contrast to the first two bioperiods, the pattern of eflow noncompliance affects rivers of all sizes. Therefore the changes in Estonian river eflows observed the rearing and growth bioperiod may present a large-scale problem for aquatic biodiversity. Especially concerning is the rapid increase in the days of noncompliance observed in 2018 for all three rivers evaluated. 

Finally, the fourth "fall spawning" bioperiod is shown in the bottom right panel of Fig. \ref{fig_plots}. Once again, the large Narva river is clearly differentiated from the remainder of the other five rivers, with the highest number of days of eflows noncompliance from 2009 to 2018. In contrast to the rearing and growth bioperiod, there is no clear large scale pattern affecting all rivers. However, there is some commonality in peaks of noncompliance in 2010 and 2015 between Narva river and Pühajõgi. We believe that the explanation for this may be geographical, as both rivers are located in northeast Estonia, and are therefore subject to similar rainfall conditions.

In summary, the results of the four bioperiods using three rivers of different size classes indicates the suitability of the proposed method for long-term water management to improve eflows in Estonian rivers. Specifically, it can be seen that the Narva river may benefit from mitigation measures, and that climate change has created a biannual rise and fall pattern of noncompliance. The information obtained through this assessment allows for water authorities to pick out both continuous and cyclical trends, and can be used to prioritize rivers by their size in order to evaluate the cumulative effects of eflows noncompliance in four different bioperiods. The next step would be to add on the ability to simulate water management actions which addressing the ecological conditions, such as the restoration of the natural flow regime, decreasing water withdrawals to agricultural and industrial end-users, etc. in order to estimate their effectiveness and improve eflow compliance \cite{methodologies}.

\section{Discussion}

Building new large-scale environmental monitoring networks and integrating the resulting data into bespoke decision-making support systems is time-consuming \cite{lovett2007}, and the additional communication costs alone can be prohibitive \cite{Liu2010}. Instead, we propose that existing OGD from national hydrological services can be repurposed for eflows compliance. The methodology is simple to apply and provides a first step in automating environmental compliance monitoring and evaluation, allowing authorities to allocate their limited resources to more important tasks such as strategic planning and enforcement \cite{Tianbao2011}.

The major contributions of this work are two-fold: First, we developed and implemented a method which allowed for rapid and scalable automation of eflows estimation and evaluation; including real-time assessment, in cases where stream flow data are available. Secondly, the method is entirely data-driven and allows for the implementation of more detailed eflow calculation methods, for example, the regionally applicable environmental method proposed recently by Paraciewicz et al. \cite{raelff}. We therefore encourage others to begin to develop and implement more advanced automated compliance systems using OGD for a wider range of applications. Furthermore, we suggest that the inclusion of compound event scenarios, which are co-occurring weather and climatic combinations of drivers and/or hazards contributing to societal or environmental risk, can be integrated into these more advanced systems to increase societal climate resilience \cite{zscheischler2020}.

We fully acknowledge that it is unrealistic to find a universal eflows method applicable to all rivers. However, we wish to point out that when data are available scenario and impact-modelling is readily achievable based solely on the available OGD. Our long-term aim is to continue developing the proposed system to automate manual functions (e.g. statistical analysis and reporting), allowing government administrators to focus on other high-priority tasks such as resilience planning \cite{Timashev2015} and the development more accurate hazard mapping for critical infrastructure \cite{Nateghi2018}.

Ultimately, IoT-based systems can provide a largely overlooked source of highly valuable OGD that enables the creation of new and improved systems for decision- and policy-making. We believe that the future of automated monitoring and regulatory compliance lies in linking OGD with the connectivity, speed and ease of use common to IoT technologies. The internet provides the ideal medium for real-time monitoring, thus helping to facilitate secure and reliable online decision-making. In our example, the modular architecture of the eflows system allows for rapid modification and optimisation of various components from functional (e.g. optimisation of routing between nodes, computing, energy use of sensing devices, and others) and nonfunctional (e.g. security and reliability of data exchange) perspectives. In addition, the integration of modules performing intelligent computing (advanced analytics, machine learning and data-driven learning) can be applied to data pipelines on different network levels (edge, fog or cloud) to detect patterns, extract and interpret the information in an operationally useful way. In future works, we will investigate the trade-offs in complexity and utility of a large-scale environmental monitoring system with multi-objective, distributed intelligence.

\section{Conclusion}
In this paper, one specific environmental regulation requirement, the EU Water Framework Directive, was studied due to the high level of uncertainty currently attached to its implementation. It was hypothesized and demonstrated that IoOGDT could be taken advantage of, combined with the EI cycle framework, and then transformed into a system that utilizes data analytics and visualization techniques to help improve the assessment, monitoring, evaluation, and reporting practices for the relevant decision-makers. In order to demonstrate this system and how it works in a real-world environment using real data, a working implementation was developed using data from the country of Estonia.

The primary advancement introduced in this work is in its transformation of a disconnected system providing a delayed response to an intelligent IoT system that provides timely awareness of the state of a given river ecosystem, allows for intelligent warnings and improved reporting to be developed, and assists decision-makers with scenario modelling, prediction, historical analysis, and reporting and monitoring requirements. Once data have been received in the cloud from the sensors, the data can be accessed and made openly available so that other systems can also utilize it and build further applications on top of such data.

The designed system dynamically links the eflow assessment with real-time monitoring in different locations. The productivity of the system in terms of improving the awareness about the state of the environment largely depends on the ability to find a trade-off between implementation complexity and application limitations. In order to use such a system for environmental compliance monitoring and evaluation, the system should interpret the results properly; otherwise, decision- and policy-making will get complicated, introducing even more uncertainties.

The research has multiple future research directions such as determination of the methods and parameters to be used for decision- and policy-making; scaling and deployment of the system to use real-time data and data from other locations; integration and validation of compliance forecasting to enable early warning; further software development activities to make the service more suitable to the target users' needs; training of the system's users, and others.

\appendices

\section*{Acknowledgment}
This research has been funding in part by Estonian Research Council grant PUT-1690 Bioinspired Flow Sensing and the Estonian IT Academy IoIT programme. We would like to thank Agne Aruväli from the Estonian Ministry of the Environment for her support, guidance and feedback on developing the "Eesti Eflows" system and School of IT, Tallinn University of Technology for providing the server. Map data copyrighted OpenStreetMap contributors and available from \url{https://www.openstreetmap.org}.

\bibliographystyle{IEEEtran}
\bibliography{IEEEabrv, bibtex/bib/bib}


\begin{IEEEbiography}
    [{\includegraphics[width=1in,height=1.25in,clip,keepaspectratio]{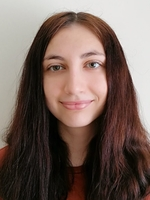}}]{Lizaveta Miasayedava} is pursuing a Ph.D. degree at the Tallinn University of Technology, Estonia. She received the M.Sc. degree in Engineering in e-Governance Technologies and Services from Tallinn University of Technology, and the B.Sc. degree in Computer Systems Engineering and Informatics from Saint-Petersburg State Electrotechnical University, Russia. She has professional experience in software engineering and web development. Her research interests include data science, applied mathematics and data-driven modelling.
\end{IEEEbiography}

\vskip -2\baselineskip plus -1fil

\begin{IEEEbiography}
    [{\includegraphics[width=1in,height=1.25in,clip,keepaspectratio]{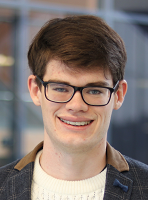}}]{Keegan McBride} received his Ph.D. in Technology Governance from Tallinn University of Technology in Tallinn, Estonia. Currently, he works as a postdoctoral researcher at the Hertie School’s Centre for Digital Governance in Berlin, Germany. His primary research interests are related to open government data, public service co-creation, complexity theory, and digital government more broadly. He has previously worked as a consultant for numerous national and international organizations on technical papers and statistical visualization projects, managed the GovAiLab and DigiGovLab at Tallinn University of Technology, and was responsible for rebuilding and managing Estonia’s open government data portal as the chief technical consultant.
\end{IEEEbiography}

\vskip -2\baselineskip plus -1fil

\begin{IEEEbiography}
    [{\includegraphics[width=1in,height=1.25in,clip,keepaspectratio]{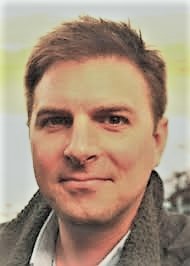}}]{Jeffrey A. Tuhtan}
    received the B.Sc. degree in civil
    engineering from California Polytechnic University,
    San Luis Obispo, CA, USA, in 2004, and the M.Sc.
    degree in water resources engineering and management
    and the Dr.Eng. degree from the University of
    Stuttgart, Germany, in 2007 and 2011, respectively.
    He is currently with the Centre for Biorobotics,
    Tallinn University of Technology, where he leads
    the Environmental Sensing and Intelligence Group.
    His current research interests include environmental
    intelligence and bio-inspired underwater sensing.
\end{IEEEbiography}

\end{document}